\documentclass[sigconf]{acmart}

\usepackage{enumitem}
\setlist[itemize]{listparindent=\parindent,align=parleft,leftmargin=0pt}
\AtBeginDocument{%
  \providecommand\BibTeX{{%
    \normalfont B\kern-0.5em{\scshape i\kern-0.25em b}\kern-0.8em\TeX}}}

\copyrightyear{2021}
\acmYear{2021}
\setcopyright{acmlicensed}\acmConference[CHI '21]{CHI Conference on Human
Factors in Computing Systems}{May 8--13, 2021}{Yokohama, Japan}
\acmBooktitle{CHI Conference on Human Factors in Computing Systems (CHI
'21), May 8--13, 2021, Yokohama, Japan}
\acmPrice{15.00}
\acmDOI{10.1145/3411764.3445159}
\acmISBN{978-1-4503-8096-6/21/05}

\begin{document}


\title[Context-Based Interface Prototyping]{Context-Based Interface Prototyping: Understanding the Effect of Prototype Representation on User Feedback}












\author{Marius Hoggenmueller}
\email{marius.hoggenmueller@sydney.edu.au}
\affiliation{Design Lab, Sydney School of Architecture, Design and Planning
  \institution{The University of Sydney}
}

\author{Martin Tomitsch}
\email{martin.tomitsch@sydney.edu.au}
\affiliation{Design Lab, Sydney School of Architecture, Design and Planning
  \institution{The University of Sydney}
}
\affiliation{CAFA Beijing Visual Art Innovation Institute, China}

\author{Luke Hespanhol}
\email{luke.hespanhol@sydney.edu.au}
\affiliation{Design Lab, Sydney School of Architecture, Design and Planning
  \institution{The University of Sydney}
}

\author{Tram Thi Minh Tran}
\email{ttra6156@uni.sydney.edu.au}
\affiliation{Design Lab, Sydney School of Architecture, Design and Planning
  \institution{The University of Sydney}
}

\author{Stewart Worrall}
\email{stewart.worrall@sydney.edu.au}
\affiliation{Australian Centre for Field Robotics
  \institution{The University of Sydney}
}

\author{Eduardo Nebot}
\email{eduardo.nebot@sydney.edu.au}
\affiliation{Australian Centre for Field Robotics
  \institution{The University of Sydney}
}

\renewcommand{\shortauthors}{Hoggenmueller and Tomitsch, et al.}

\begin{abstract}
The rise of autonomous systems in cities, such as automated vehicles (AVs), requires new approaches for prototyping and evaluating how people interact with those systems through context-based user interfaces, such as external human-machine interfaces (eHMIs). In this paper, we present a comparative study of three prototype representations (real-world VR, computer-generated VR, real-world video) of an eHMI in a mixed-methods study with 42 participants. Quantitative results show that while the real-world VR representation results in higher sense of presence, no significant differences in user experience and trust towards the AV itself were found. However, interview data shows that participants focused on different experiential and perceptual aspects in each of the prototype representations. These differences are linked to spatial awareness and perceived realism of the AV behaviour and its context, affecting in turn how participants assess trust and the eHMI. The paper offers guidelines for prototyping and evaluating context-based interfaces through simulations.
\end{abstract}




\maketitle

\section{Introduction}

With the rise of autonomous systems and their application in everyday products, the human-computer interaction (HCI) community has turned its attention towards developing ways for supporting the design of such systems. Within the context of cities, autonomous systems promise to transform urban mobility and to automate services \cite{Macrorie2019}. Recent trials of automated vehicles (AVs) as early protagonists of autonomous systems in cities, have primarily focused on making the technology work. 
However, a key for the successful uptake of AVs is the careful consideration of trust, usability and user experience, as found in a study on automated driving \cite{Frison2019}. 
Within an urban environment, this extends to the design of the external human-machine interface (eHMI) that AVs use to communicate their internal state and their intent to pedestrians \cite{Dey2020taming, Loecken2019, Owensby2018}. 

Prototyping and evaluating eHMIs with prospective users in urban environments is extremely challenging, as it is associated with high costs of real-world prototypes (e.g. a self-driving car) and potential risks to participants. To address these challenges, HCI researchers have turned to using various simulation platforms and prototype representations that allow them to simulate eHMI concepts in a lab environment \cite{Shuchisnigdha2017, Colley2019}. This includes the use of video recordings to study pedestrian interactions with an eHMI \cite{Hollaender2020, Song2018} and computer-generated (CG) prototypes in a virtual environment (VR) to evaluate how pedestrians would cross in front of an AV equipped with an eHMI \cite{Chong2013}. 

Previous simulation studies have primarily focused on prototyping and evaluating specific interface concepts (e.g. \cite{Gerber2019}), assessing how participants experience the simulation (e.g. \cite{Flohr2020}) and comparing the sense of presence across CG and real-world representations (e.g. \cite{Dohyeon2020}). To our knowledge, no studies have been carried out to date to investigate in what ways different prototype representations affect how participants provide feedback on the prototype itself. To address this gap, we implemented a mixed-methods study in which we compared three prototype representations: real-world VR, CG VR and real-world video. As a case study, we chose a ride-sharing scenario (captured from the perspective of a pedestrian waiting for their vehicle to arrive) in a shared urban environment, where pedestrians, cyclists and maintenance vehicles share the same road. We chose this scenario as previous research has found that people consider interactions with AVs more important in shared environments \cite{Schieben2019}. Our research team involved interaction designers, urbanists and engineers, which allowed us to take a holistic approach to designing the AV prototype and the scenario used in our study. To that end, we used a fully functional AV that was specifically designed for a shared environment and equipped with an eHMI in the form of a low-resolution display. 

The paper makes three contributions to the field within HCI that is concerned with the design of human-machine interfaces for autonomous systems. (1) It presents the first comparative study of different simulation approaches for evaluating eHMIs from a pedestrian perspective. (2) It provides empirically based insights on what participants focus on when assessing trust and user experience across real-world VR, CG VR and real-world video prototype representations. (3) It offers guidelines for how to create context-based interface prototypes for lab-based evaluation studies. 

\section{Related Work}

Within the broader context of autonomous systems, this paper specifically draws on and contributes to (1) the design of eHMIs, (2) prototyping approaches and the simulation of interactions between people and AVs, and (3) studies of simulation platforms. 

\subsection{External Human-Machine Interfaces}

Being designed to communicate the system's awareness and intent, eHMI concepts include projection-based solutions \cite{Nguyen2019} and displays attached to the vehicle \cite{Eisma2020}, thereby supporting various communication modalities, from abstract \cite{Dey2020} to symbolic \cite{Hollaender2019} to textual \cite{Bazilinskyy2019}. The study reported in this paper contributes to this field through the systematic investigation of the various prototyping representations that are currently available to evaluate eHMIs and other context-based interfaces for autonomous systems (e.g. mobile robots, drones) in complex urban environments.

\subsection{Prototyping and Simulation}

The creation of prototypes is an integral part of a human-centred design process \cite{Buxton2007} and can fulfil various purposes; for example, prototypes are often used to evaluate certain aspects of a design with users, before further development stages commence \cite{Buchenau2000}. Lim et al. \cite{Lim2008} highlight the importance of understanding the fundamental characteristics of prototypes and the careful selection of representational forms, prototyping materials and resolutions, as these influence the judgement of a target design concept.

Given the complexity, cost and potential risk to participants,  
associated with designing and evaluating interfaces for and interactions with autonomous systems, researchers have turned to a wide range of methods and techniques, such as Wizard of Oz, video and simulation prototyping \cite{Pettersson2017}. 
In particular, CG VR has been found to be a promising approach for simulating autonomous systems and their interfaces in a safe environment \cite{Nascimento2019TheRO}. CG VR allows for assessing the user experience (UX) of an interaction in a contextual environment \cite{Rebelo2012} while increasing controllability and reproducibility \cite{Winter2012}. Research on pedestrian safety has further demonstrated that participant behaviour in CG VR matches real-world norms and that participants found the VR environment to be realistic and engaging \cite{Shuchisnigdha2017, Mahadevan2019}. Simulations also have the advantage of allowing for rapid prototyping approaches, as various interface elements can be quickly exchanged and evaluated \cite{Alvarez2015, Gerber2019}. Simulation studies are not limited to CG environments with some studies employing video \cite{Hollaender2020, Song2018} or 360-degree video based VR \cite{Burns2020, Dohyeon2020, Gerber2019, Velasco2019} as a way to simulate the experience of interacting with real-world prototypes.

When it comes to the evaluation of context-based interfaces, it is important that the simulated environment offers realistic experiences and provokes similar user behaviours to those observed in the real world. 
Here, results from similar HCI research domains are promising; for example, M{\ae}kel{\ae} et al. have reported that in the area of public display research, they were able to observe similar user behaviour in virtual compared to real-world settings \cite{Maekela2020}. They therefore propose virtual field studies as an alternative to real-world studies, offering similar ecological validity but at a reduced effort. 

A key purpose of prototypes is to collect feedback from prospective users. To that end, Pettersson et al. \cite{Pettersson2019} found that the overall user experience was similar when comparing in-vehicle systems in VR and in the field, but that participants provided less feedback in VR. They further observed that users had difficulties to separate judgements about the evaluated prototype and the system through which the prototype is presented. Similar findings were reported by Voit et al. for the evaluation of smart artefacts \cite{Voit2019}; besides differences in reported feedback, they also found that evaluation methods can influence study results. This paper sheds further light on how user feedback varies across different prototype representations.

\subsection{Simulation Platforms}
Regardless of the simulation platform being used, i.e. CAVE-like setups \cite{Flohr2020}, screen-based driving simulators or VR headsets, a major consideration in the development of simulator platforms is to offer users a high sense of presence \cite{Lessiter2001}. This can be achieved through various measures, such as increasing interaction fidelity \cite{Rogers2019}, motion fidelity \cite{Dohyeon2020} and offering a high visual realism \cite{Gisbergen2019}. Previous research has shown that higher visual realism enhances realistic response in an immersive environment \cite{Slater2009}. 

To that end, 360-degree video is a promising alternative to CG, as it results in higher perceived fidelity and presence compared to CG simulations \cite{Dohyeon2020}. In addition to the higher visual realism, users' sense of presence also benefits from the familiarity with the environment when using immersive real-world videos \cite{Gerber2019}.
Importantly, real-world video is able to represent not only the prototype but also the context at a high level of fidelity \cite{Flohr2020}, which is comprised of various elements, such as audio-visual impressions, the physical environment and the presence of other people and the user’s relationship with them \cite{Kray2007}. These elements might influence how participants experience an eHMI in a simulated situation \cite{Krome2015}.\\

The recent uptake in research on prototyping strategies for eHMIs within the HCI community points to VR and 360-degree video simulations as competing emerging trends. Yet, despite the many promising concepts, the complexity of the context under investigation means that more work is required to further understand the inherent qualities of those prototyping representations. Previous work has highlighted the importance of understanding the fundamental characteristics of physical prototypes in the context of interactive products \cite{Lim2008}. To the best of our knowledge, characteristics of emerging prototype representations, such as 360-degree video simulations and VR, in relation to increasingly important outcome variables, such as system trust, have not yet been systematically studied. In particular, a systematic evaluation of the effect of different prototype representations on user feedback – and therefore study results – is still lacking. This study represents a first attempt to address this gap, which we argue will not only inform research on and the design of AVs, but also other categories of urban technologies and autonomous systems, such as robotic interfaces \cite{Tomitsch2021} and pulverised displays \cite{hoggenmueller2019self}.

\section{Evaluation Study}

Building on previous work and to address the gap identified in the review of previous studies, we set out to investigate how user feedback varies across different prototype representations. Rather than evaluating a specific eHMI, our aim was to understand the factors that influence user feedback on eHMIs. This aim follows the trajectory from early work in HCI that reported on differences in user feedback when evaluating paper versus interactive prototypes \cite{Walker2002, Liu2003}. As previous studies of VR simulations have found sense of presence to be an important factor, we formulated our first research question (RQ1) to measure sense of presence for each of the prototype representations: \emph{How does the prototype representation affect user’s sense of presence?}. 

The subsequent two research questions that drove our study design were formulated to measure specific user feedback sought when evaluating human-machine interfaces, with previous studies highlighting trust and UX as important aspects \cite{Kaur2018, Lacher2014, Frison2019}. Thus, the second research question (RQ2) was \emph{How does the prototype representation affect perceived user’s trust in the eHMI?} and the third question (RQ3) was \emph{How does the prototype representation affect the perceived UX of the eHMI?}. 

\subsection{Study Design}

We adopted a between-subject approach for the gathering of quantitative data to assess sense of presence, trust and user experience, thus reducing learning effects and avoiding carryover effects from repeated measures. To that end, we balanced the distribution of participants across the three prototype representations. After experiencing the assigned prototype representation, participants were asked to complete a set of questionnaires and to partake in a semi-structured interview. This was followed by participants experiencing the same scenario in the remaining two prototype representations. At the conclusion of the study, participants took part in a second semi-structured interview. This approach was chosen to allow participants to compare their perceived sense of trust and UX across all three representations.  

\subsection{Prototype Representations}

We opted to compare real-world VR (referred to as RW-VR), computer-generated VR (CG-VR) and real-world video (RW-Video), hence adopting two simulation platforms (VR and video). RW-VR is increasingly used in simulation studies given that 360-degree cameras are becoming more affordable and widely available \cite{Jokela2019}, and due to the higher level of fidelity of real-world video \cite{Flohr2020}. CG-VR is a commonly used representation in pedestrian-AV safety research (e.g. \cite{Shuchisnigdha2017, Nascimento2019TheRO}). RW-Video was included as video prototypes can be useful when evaluating context-based interfaces online \cite{Dey2020}. Video prototypes are further less complex and lower-cost in terms of the evaluation setup.

The eHMI and the trajectories were the same for all three prototype representations in terms of depicted eHMI hardware (i.e. resolution; display technology), displayed content (i.e. light patterns) and context (i.e. location and time when a specific vehicle behaviour and light pattern was triggered). Differences would only occur due to the inherent nature of the prototype representation, whose effects on study results were part of the investigation.

\subsubsection{RW-VR}
For creating the RW-VR prototype representation, we worked in close collaboration with researchers from urbanism and from the engineering department of our university. We used a fully functional AV developed by the engineering department as a cooperative autonomous electric vehicle (CAV) platform \cite{Baber2005} with hardware designed by AEV Robotics\footnote{\url{https://aevrobotics.com/}, last accessed September 2020}. The vehicles - being small, efficient and electrically powered - were designed for the purpose to operate safely in low speed road environments (under 40kph) and in shared environments where the vehicles would be operating in close proximity to pedestrians \cite{Pavone2015, Spieser2014}. The platforms have the sensing and computation capacity to eventually operate at level 5 as defined by the Society for Automotive Engineers (SAE) for autonomous driving. The system is based on the robot operating system (ROS) which is a middleware for robotic platforms that enables and promotes modular system design.

\begin{figure}
  \centering
  \includegraphics[width=\linewidth]{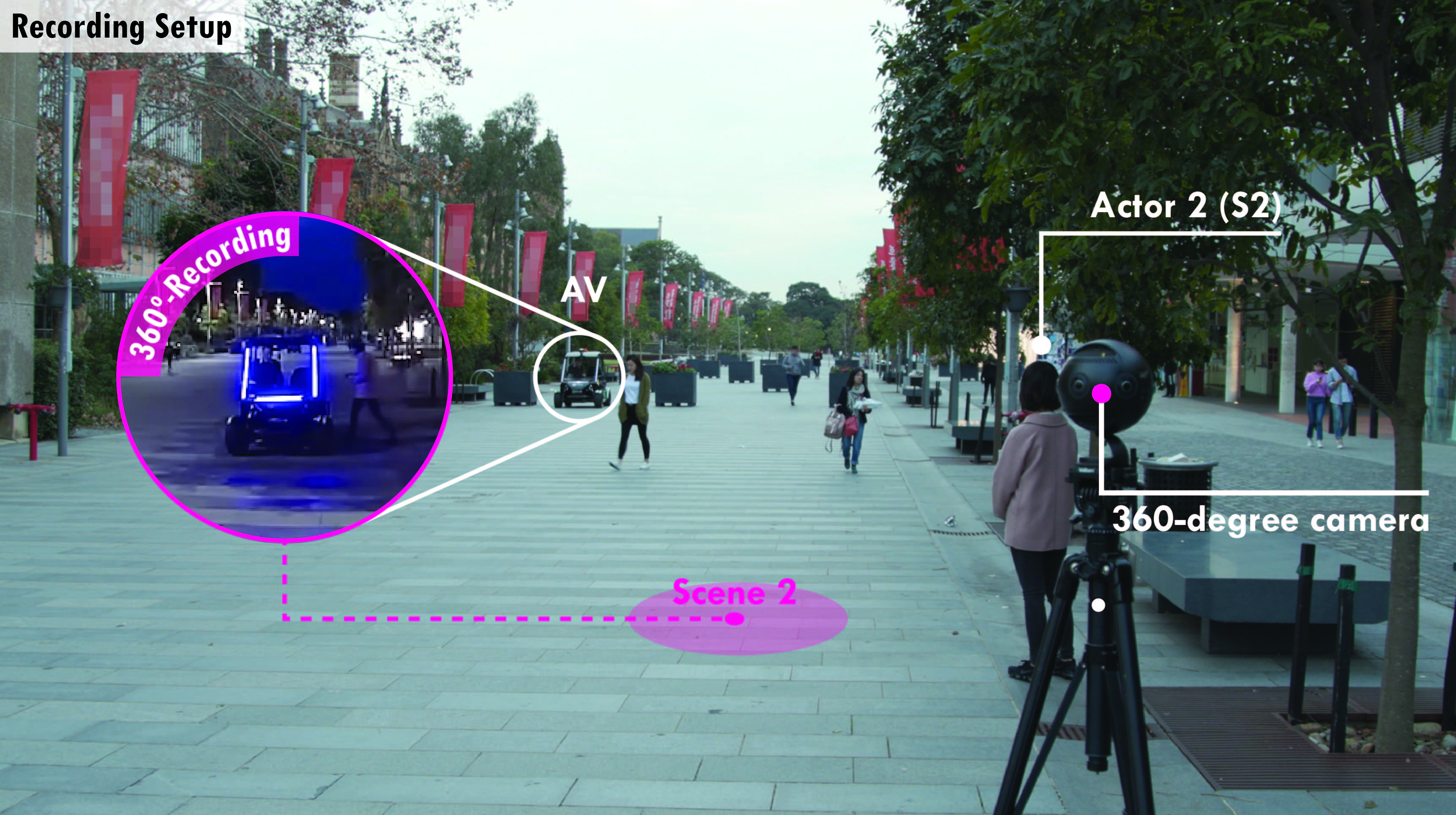}
  \caption{Recording setup for the immersive 360-degree real-world video prototype representation.}
  \Description{Figure 1 shows an annotated image of the recording setup, including the 360-degree camera setup, an actor and the approaching AV in the back.}
  \label{fig:recording_setup}
\end{figure}

For the purpose of this study, we designed a low-resolution (low-res) lighting display functioning as an eHMI to communicate the shared AV's intent and awareness, as well as enabling users to identify their car, following recommendations from previous studies \cite{Owensby2018, Boeckle2017}. The display consisted of LED strips installed on three sides of the front window as shown in Figure~\ref{fig:recording_setup}. The LED strips featured a pitch of 60 pixels per meter, resulting in a total of 145 LEDs. 
To improve the viewing angle and to create the illusion of a light bar (rather than a distinct set of point light sources), a diffuser tube of opal white acrylic was added. The LEDs were controlled via an Arduino board, which was connected to the system of the vehicle. A python ROS node was constructed that read the information from the vehicle state by subscribing to the relevant information. All light patterns were triggered in real-time based on the sensed information (awareness) and the state of the AV platform (intent).

\begin{figure}
  \centering
  \includegraphics[width=1\linewidth]{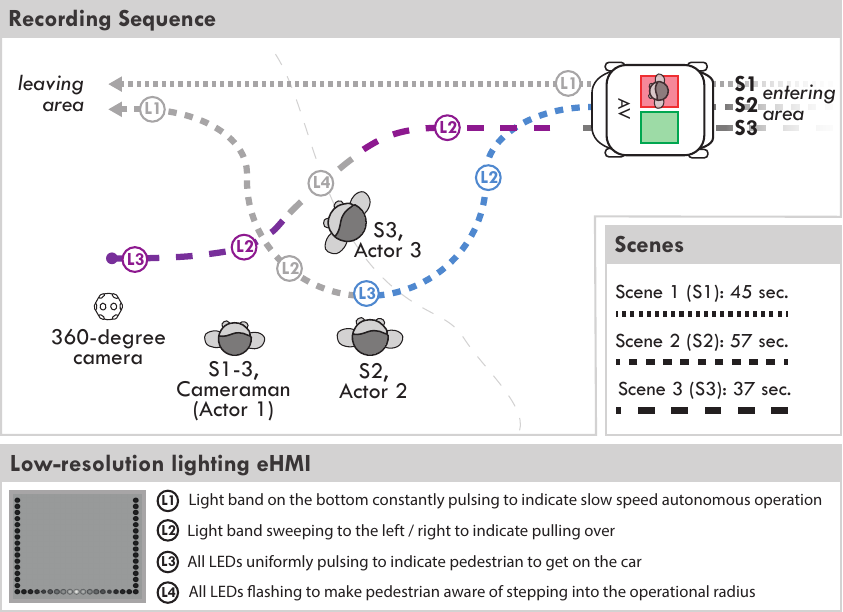}
  \caption{Scenes (S1-3) and trajectories for the 360-degree video recording and the light patterns (L1-L4) used in the AV's eHMI for the various simulated interactions. Colours in the trajectories represent the colour encoding, which has been used for riders to identify their vehicle (i.e. blue for Actor 2, purple for the user experiencing the prototype in VR).}
  \Description{Figure 2 shows an illustration of the recording plan with all the scenes and implemented light patterns.}
  \label{fig:recording_plan}
\end{figure}

We developed a set of light patterns to demonstrate the usage of an eHMI interface for shared AV services following a user-centred design process supported through a purpose-built prototyping toolkit (involving workshops with 14 experts) \cite{Hoggenmueller2020ozchi}. We do not dive deeper into the design of the eHMI itself here, as this is not the focus of the contributions reported in this paper. The final sequence of light patterns along with the scenes used in the prototype representation are depicted in Figure~\ref{fig:recording_plan}. In total, we recorded three scenes to demonstrate the eHMI interface in a shared AV scenario. We staged and recorded the scenes in a shared environment (one of our university's main avenues) with an Insta360 Pro 2\footnote{\url{https://www.insta360.com/product/insta360-pro}, last accessed September 2020} camera, which can record 360-degree panorama videos in 8K 3D. The scenes (represented from the perspective of the study participant) included: (1) The AV passing through the shared environment without any staged interactions with pedestrians. (2) The AV pulling over and picking up another pedestrian (Actor 2 in Figure~\ref{fig:recording_plan}). (3) The AV indicating to pull over to the camera stand. In this trajectory, another pedestrian (Actor 3 in Figure~\ref{fig:recording_plan}) forces the AV to slow down and stop, demonstrating how a pedestrian safely crosses in front of the AV. An additional person was placed directly behind the camera in all three scenes (Actor 1 in Figure~\ref{fig:recording_plan}), giving the appearance of another rider waiting for their own shared AVs. This was to constraint participants' movement in the simulation, as 360-degree video does not allow for motion when imported into VR. 

All three scenes were recorded with the same AV and therefore recorded consecutively. The vehicle was operating based on pre-computed trajectories that mimicked the desired vehicle behaviour for the purpose of recording the 360-degree video. The vehicle was operating a `virtual bumper' which is a system that detects obstacles in (or adjacent to) the proposed vehicle trajectory and reduces the speed based on a time-to-collision calculation. Due to safety regulations, a licensed operator had to sit in the AV – in case of having to manually bring the AV to a halt. However, for the purpose of the recordings, we were able to remove the steering wheel, thus conveying clearly to participants that the car was driving autonomously. 

After recording the scenes with the 360-degree camera, we used Adobe Premiere and Adobe After Effects for post-processing. As we recorded the scenes in early evening hours for better visibility of the low-res lighting display, we had to apply the Neat Video\footnote{\url{https://www.neatvideo.com/}, last accessed September 2020} filter to reduce image noise, while still preserving fine details, such as people's faces. We then combined the three scenes, added a short blend transition between them, and exported them into a single 3D over-under video file. To experience the stereoscopic 3D 360-degree video with a VR headset (HTC Vive), we imported the video file into Unity and applied it as a render texture on a skybox material. To convey the immersive audio recording of the scene soundscape and increase a sense of presence, we used stereo headphones.

\subsubsection{CG-VR}
To create the same three scenes from the RW-VR in the CG-VR, we commissioned a 3D simulation designer with a background in interaction design and a specialisation in 3D modelling and creating immersive virtual products with more than 8 years of professional experience. We provided an overview of the scene recordings (similar to Figure \ref{fig:recording_plan}), the 360-degree video from the RW-VR as a reference as well as a building information model (BIM) of the university campus avenue. The designer further conducted several visits to the physical site to better assess the dimensions and proportions of the shared space environment and the surrounding buildings. For the design of the AV, we provided the 3D designer with technical drawings, photographs and videos of the actual AV. The car model was created in Autodesk 3ds Max, using emissive materials for the low-res lighting display in order to replicate the lighting effects and aesthetics as realistically as possible.The car model and the low-res lighting display were then animated in Unity. We deliberately decided against using an existing autonomous driving simulator with a sensor suite (e.g. Carla \cite{Dosovitskiy2017}), as Unity has been used for the majority of eHMI research and provides more flexibility for designing and prototyping customised context-based interfaces and the surrounding environment. For creating the actors and surrounding pedestrians from the 360-degree video, models from a library providing 3D scanned people\footnote{\url{https://renderpeople.com/}, last accessed September 2020}, were used and customised for our scenario. Throughout the design process, we arranged several meetings with the 3D designer and also tested the prototype in VR. Through this iterative approach, changes were made to the atmospheric lighting of the scene and interactions of pedestrians with the AV were adjusted to match the details from the 360-degree video. 
For the experiment we used the same VR headset as for the RW-VR simulation. We imported the immersive audio recording to reduce any effects of sound as a potentially confounding variable. 

\begin{table}[b]
  \caption{Number, gender, age and previous VR experience of participants for each prototype representation.}
  \label{tab:participants}
  \begin{tabular}{lccc}
    \toprule
    &\textbf{RW-VR}&\textbf{CG-VR}&\textbf{RW-Video}\\
    \midrule
    \vspace{0.1cm}
    \textbf{n (m/f)} & 14 (7/7) & 14 (7/7) & 14 (8/6)\\
    \textbf{Age} & M=31.42 & M=32.57 & M=32.42\\
    \vspace{0.1cm}
    & SD=8.6 & SD=10.38 & SD=8.48\\
    \textbf{Prev. VR-Exp.}\\
    never & 3 & 3 & 3\\
    less than 5 times & 8 & 9 & 6\\
    more than 5 times & 3 & 2 & 5\\
  \bottomrule
\end{tabular}
\end{table}

\subsubsection{RW-Video}
For the real-world video prototype representation, we used the previously recorded and post-processed 360-degree video as a source. We used Adobe After Effects to map the equirectangular into a 2D rectilinear video projection. In order to highlight the first-person nature of the experience (as opposed to having the participant just passively watching the video as a passer-by in the environment), we animated the viewing angle of the video to ensure that the AV was always in the centre of the image. Thus, if the AV was driving out of the scene, the camera would follow its trajectory as if the participant were waiting for their own car. We exported the final video as a 1080p resolution, 16 to 9 video file. For the experiment, we displayed the video in full-screen mode on a 24-inch monitor. As in the VR prototype representations, we also used the same stereo headphones to convey the immersive audio soundscape.

\subsection{Participants}
We recruited 42 participants (22 male, 20 female) between the ages of 21 and 57 (\textit{M=32.05}, \textit{SD=9.13}). Participants were recruited from our university's mailing lists, flyers and social networks; all participants voluntarily took part in the experiment and initial contact had to be made by them, following the study protocol approved by our university's human research ethics committee. Participants were randomly assigned to one of the three conditions to start with; the two remaining conditions that participants experienced before the post-study interview were counterbalanced. Further, we balanced participants' age, gender and previous experience in VR across the three prototype representations with the help of an online screening questionnaire that we sent to participants prior to the experiment. Table~\ref{tab:participants} shows participant characteristics for each of the conditions that they experienced first.

\begin{figure*}
  \centering
  \includegraphics[width=1\textwidth]{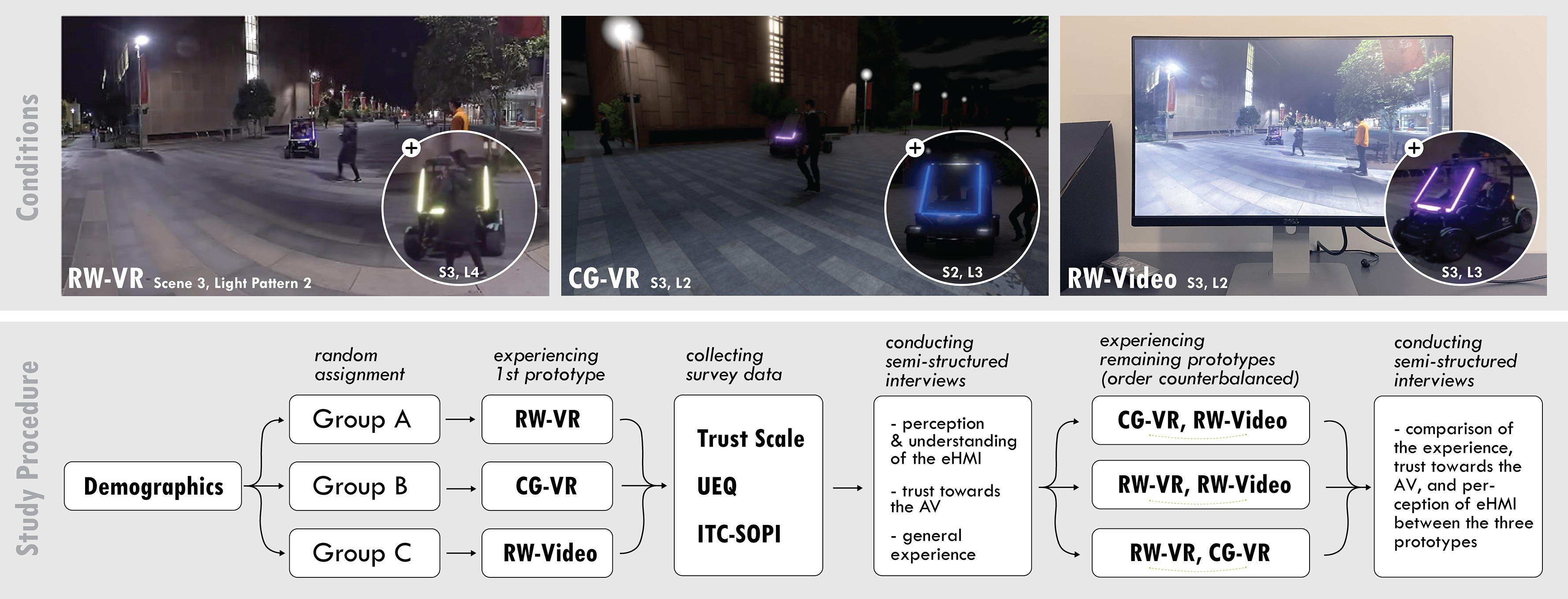}
  \caption{Frames from the RW-VR and CG-VR simulations and the RW-Video setup used as conditions in the study (top). Study procedure showing the way participants experienced each of the prototype representations and the data collected (bottom).}
  \Description{Figure 3 shows of all three prototype representations investigated in the user study, and an overview of the study procedure.}
  \label{fig:procedure}
\end{figure*}

\subsection{Study Procedure}
Upon arriving in our lab, participants received a short introduction about the research topic on shared AVs and eHMIs. We informed participants that the aim of our study was to evaluate trust and UX for eHMIs in a shared AV scenario. We did not mention the comparison of representations in order to avoid biases towards the questionnaire and the interview that we conducted after the first experienced prototype representation. We then asked participants to fill out the consent form to take part in the study, followed by a short questionnaire to collect data on demographics. Before experiencing the first prototype representation, we shortly briefed participants about the scenario they would experience, following advice from previous work suggesting that providing users with a meaningful narrative context increases their inner presence \cite{Gorini2011}. To further immerse them into the scenario of waiting for a shared AV, we presented them with a mock-up interface on a mobile phone. The interface followed the layout of existing ride-sharing services and displayed: (a) a map of the location where the participants were supposed to wait for their vehicle, (b) the vehicle's current position approximately 2 minutes away from the participant, (c) the colour which was assigned by the system for the participant to recognise their vehicle (in this case purple), and (d) the mock user profile of the person whom they would share the vehicle with. 

After experiencing the first prototype representation, participants were asked to complete a set of standardised questionnaires, which took between 9 to 13 minutes. In a next step, we conducted a semi-structured interview (M=7min 43sec, SD=2min 54sec).
After consecutively experiencing the two remaining prototype representations, we conducted a semi-structured post-study interview (M=9min 34sec, SD=2min 53sec).

The duration of each experienced scenario was 2 minutes and 19 seconds (same duration for each condition). 
We chose this time frame carefully based on initial tests within the team, ensuring that there was sufficient time for participants to get familiar with the context and to adjust to the immersive experience, yet short enough to avoid fatigue. 
The whole study took approximately 45 minutes for participants to complete. We informed participants that they could stop the experiment at any time, for example, should they experience motion sickness, however, none of the participants had to stop the experiment. The conditions and study procedure are illustrated in Figure~\ref{fig:procedure}.

\subsection{Data Collection}
Throughout the experiment we collected both quantitative and qualitative data, following a mixed-methods approach \cite{Creswell2014}.
In the following we provide an overview of our data collection. We present the questionnaires in the same order as participants were asked to complete them during the experiment.

\subsubsection{Questionnaires}
In order to measure participants' subjective perception of trust towards the AV, we used a standardised trust scale that was designed for the measurement of trust in autonomous systems \cite{Jian2000}. The questionnaire which has been widely used in the context of research on autonomous vehicles \cite{Pettersson2019, Frison2019} consists of two subscales to calculate an overall \textit{trust} score (7 items) and an overall \textit{distrust} score (5 items); all items correspond to 7-point Likert scales. We instructed participants to assess trust by considering the AV as a single system and based on how they experienced the AV in the presented scenario. 

To assess participants' UX of the eHMI, we used the UEQ questionnaire \cite{Laugwitz2008}. The questionnaire consists of 26 bipolar items (7-stage scale from -3 to +3) to calculate 6 UEQ subscales: \textit{attractiveness} (overall impression of the product), \textit{perspicuity} (how easy it is to get familiar with the product), \textit{efficiency} (solving tasks without unnecessary effort), \textit{dependability} (feeling in control), \textit{stimulation} (how exciting and motivating it is to use the product), and \textit{novelty} (how innovative and creative the product is). For the UX questionnaire, we instructed participants to consider the low-res lighting interface of the AV as experienced in the presented scenario. 

To assess participants' media experience and sense of presence in the three prototype representations, we employed the ITC-Sense of Presence Inventory (ITC-SOPI) \cite{Lessiter2001}. The questionnaire is well established to compare sense of presence across a wide range of media systems, and has been previously used for comparing semi-autonomous driving systems in VR and in the field \cite{Pettersson2019}. The questionnaire consists of 38-items (5-point Likert) to calculate 4 subscales: \textit{spatial presence} (assessing the sensation of being in a displayed environment), \textit{engagement} (measuring the intensity of the experience and feeling of being involved), \textit{ecological validity} (naturalism of the displayed environment and sensation that displayed objects are solid), \textit{negative effects} (assessing potential negative effects such as motion sickness).

\subsubsection{Interviews}
We collected qualitative data in the form of semi-structured interviews.
In the first round of interviews, conducted after participants experienced the first prototype representation and following the questionnaires, we asked questions about (1) understanding of the light patterns, (2) trust towards the vehicle and (3) comments on the experience. In the post-study interview, conducted after participants experienced the remaining two presentations, we asked questions about (1) differences between the three prototype representations in terms of their experience, (2) whether experiencing the remaining two presentations changed participants' perceived trust towards the AV and (3) perception and understanding of the lighting display.

\subsection{Data Analysis}

\subsubsection{Questionnaires}
We first conducted a descriptive analysis of our questionnaire data to obtain an overview about the relationship between each predictor and the outcome domain variable. Thus, we calculated means and standard deviations after an internal reliability assessment of the scales calculating Cronbach's alpha. Overall internal reliability was excellent for both \textit{trust} subscales ($\alpha$ >= 0.9). For the UEQ questionnaire, item reliability was acceptable for \textit{efficiency}, \textit{stimulation} and \textit{novelty} ($\alpha$ > 0.7), and good for \textit{attractiveness}, \textit{perspicuity} and \textit{efficiency} ($\alpha$ > 0.8). For the ICT-SOPI, overall internal reliability was excellent for \textit{spatial presence} and \textit{engagement} ($\alpha$ > 0.9), good for \textit{negative effects} ($\alpha$ = 0.83), and acceptable for \textit{ecological validity} ($\alpha$ = 0.71). 

We conducted a univariate analysis of variance (ANOVA) for each outcome domain of the questionnaires. We used side-by-side box plots to assess if the data was approximately normally distributed. In case of normal distribution, we calculated one-way ANOVA, otherwise the Kruskal-Wallis rank sum test was utilised. In case of significant differences, we performed post-hoc tests using Benjamin-Hochberg (BH)-corrected p-values. 

\subsubsection{Interviews}

All interviews were transcribed by a professional transcription service. Two coders worked collaboratively to analyse the data from both interviews. However, each coder started the analysis with a different set of interviews and independently developed the codebook. We reviewed each other's codebooks afterwards to discuss the difference and made adjustments where needed. 

The data from the post-study interview was used to assess participants' preferences and to identify the reasons for those preferences as well as for changes in terms of their perceived trust towards the AV. These data were analysed following a deductive thematic analysis approach \cite{BraCla06}, using a digital whiteboard with sticky notes. The identified themes were used to structure the Discussion section. To further illustrate specific observations around the identified themes, relevant quotes were selected from the first interview. 

The data from the first interview was used to assess perceived trust towards AV and user experience of the lighting system. Using the same analysis approach with the post-study interview, we identified key aspects of trust and user experience that changed with prototype representations. Explanation of these changes, however, were found mainly in the analysis of the post-study interview.

\section{Results}

\subsection{Sense of Presence (RQ1)}

\begin{table}
  \caption{Mean and standard deviations for ITC-SOPI ratings for the three prototype representations (max=5, min=1).}
  \label{tab:itc}
  \begin{tabular}{lcccc}
    \toprule
    &\textbf{RW-VR}&\textbf{CG-VR}&\textbf{RW-Video}&\textit{p-value}\\
    &(M / SD)&(M / SD)&(M / SD)\\
    \midrule
    \vspace{0.2cm}
    \textbf{Spatial Pres.} & \textbf{3.15} / 0.56 & \textbf{3.21} / 0.43 & \textbf{2.32} / 0.96 & <0.01\\
    \vspace{0.2cm}
    \textbf{Engagement} & \textbf{3.93} / 0.60 & \textbf{3.91} / 0.52 & \textbf{2.84} / 0.64 & <0.001\\
    \vspace{0.2cm}
    \textbf{Ecol. Val.} & \textbf{4.4} / 0.50 & \textbf{3.84} / 0.60 & \textbf{3.74} / 0.59 & <0.01\\
    \textbf{Neg. Effects} & \textbf{1.82} / 1.00 & \textbf{1.48} / 0.40 & \textbf{1.52} / 0.44 & 0.369\\

  \bottomrule
\end{tabular}
\end{table}

\subsubsection{ITC-SOPI}
Results of the ITC-SOPI (see Table~\ref{tab:itc}) show above middle rating sense of presence for RW-VR and CG-VR, and below middle rating for RW-Video. Engagement ratings are high for RW-VR and CG-VR, and slightly above middle rating for RW-Video. Ecological validity scale is high for CG-VR and RW-Video and very high for RW-VR. Negative effects are low for all three prototype representations, with slightly higher ratings for RW-VR. Univariate ANOVA found no significant main effect of prototype representation on \textit{negative effects} (\textit{F(2, 37) = 1.023, p = 0.369}). However, a significant main effect of prototype representation was found for \textit{spatial presence} (\textit{F(2, 39) = 7.258, p < 0.01}), \textit{engagement} (\textit{F(2, 39) = 15.77, p < 0.001}) and \textit{ecological validity} (\textit{F(2, 39) = 5.424, p < 0.01}). For \textit{spatial presence}, post-hoc tests revealed significant differences between RW-VR and RW-Video, as well as CG-VR and RW-Video (both with \textit{p < 0.01}). For the \textit{engagement} scale, post-hoc tests revealed significant differences between RW-VR and RW-Video, as well as CG-VR and RW-Video (both with \textit{p < 0.001}). Finally, for \textit{ecological validity}, post-hoc tests revealed significant differences between RW-VR and CG-VR, and also RW-VR and RW-Video (both with \textit{p < 0.05}).

\subsubsection{Qualitative Feedback} 

In terms of the media experience (i.e. how the prototype was presented), the post-study interviews showed that RW-VR was favoured by the majority of participants (n=30), followed by CG-VR (n=5) and RW-Video (n=1). There were 4 participants who favoured both VR representations and 2 participants did not have a preference. Being the least immersive, the RW-Video was perceived by many participants as \textit{`boring’} (P10, P28, P37). Most participants felt like they were \textit{`watching’} a video (n=9) and were not really \textit{`being there’} in the scene (n=8). As P34 stated, \textit{`you are not present at that particular place, so you are distancing yourself from the actual situation’}. Henceforward, we discuss two predominant themes from our thematic analysis in more detail:

\begin{itemize}[label={}]

\item (1) Visual realism: Between the two VR simulations, participants commented positively on RW-VR due to the higher realism of the presented environment. The RW-VR prototype representation allowed participants to \textit{`naturally [...] step in that environment'} (P8) and \textit{`not [get] distracted by the novelty of being in a virtual world'} (P24). The real-world environment made it easy for participants to quickly understand the scenario, which P18 related to a perceived reduction of cognitive load: \textit{`[...] because it is so much more realistic, your brain doesn't have to do the work to try to create the picture and make sense of it, which means you can actually focus on the aspects of the car and what it does and the way it communicates’}. Participants who experienced the RW-VR also mentioned in the first interviews that they were impressed by the high realism (n=6). Two participants stated in this regard that they \textit{`haven't experienced something that well put together in VR'} (P5) and were rather expecting a representation that is \textit{`cartoon’}-like (P5) or \textit{`game'}-like (P21). Thirteen participants explicitly stated that the CG-VR felt \textit{`more like a game'} and that it \textit {`seemed weird and somehow detached from reality'} (P27). The subjective experience of being present was not as strong, as reported by P33: \textit{`I felt like I was injected into a scene'}.
Some participants mentioned that they were \textit{`distracted by'} (P18, P24, P33) or \textit{`focused on'} (P15, P26) the imperfections in the simulated environment: \textit{`I was thinking a lot about how this computer world was created. I was just looking at the patterns on the trees and looking at the movement of [people]'} (P15). P7 mentioned an interesting aspect about feeling related to other people within an immersive scene: \textit{`it’s easier to relate to an image of a real human than to an avatar'}. For the CG-VR she would have expected to \textit{`see [her] hands like an avatar hand as well'}, so she could see herself as one of the people there and connect with them.

\item (2) Interaction fidelity: In the RW-VR prototype representation, participants were able to look around the environment but could not move around as naturally as in CG-VR. Motion sickness might be experienced by those who tried to walk a few steps, as pointed out by P18: \textit{`If you move, but the picture does not move accordingly, your brain will make you sick’}. Six participants noticed the nature of 360-degree video and did not attempt to move: \textit{`I felt like I was on a fixed camera stand’} (P37). Five participants said that they did not feel it was possible to interact, because things were \textit{`too realistic'} (P16) or had \textit{`already been happening’} (P29). Despite the difference in terms of interactivity, the number of participants who felt the urge to respond physically to the AV was the same for both RW-VR (n=5) and CG-VR (n=5). P31 asked the researcher if she could walk to the vehicle and \textit{`actually sit there'}. P35 raised a similar point: \textit{`when it arrived for me I would've liked to walk up to it’}, expressing the desire to explore the experience holistically, including to commute in a shared AV. In the CG-VR prototype representation, participants (n=6) thought that interaction was possible since things were \textit{`rendered’} and \textit{`it can take other inputs'} (P37). Two participants suggested the potential impact of interactivity on feelings of immersion. P34, for example, said: \textit{`if there was a possibility of interaction […] like going in front of the car and it stopped, then I would say that the second prototype [CG-VR] will be much more immersive than the third one [RW-VR]’}.

\end{itemize}

\subsection{Trust (RQ2)}

\subsubsection{Trust Scale}
Descriptive data analysis of the subjective trust ratings \cite{Jian2000} show that participants' trust towards the AV was higher for the VR representations than for the non-immersive RW-Video, with highest trust in RW-VR (see Table~\ref{tab:trust}). Conversely, participants' distrust towards the AV was higher for RW-Video than for RW-VR or CG-VR, with lowest distrust in the RW-VR. Yet, no statistically significant difference could be found. That said, RW-VR had the lowest standard variations for both trust and distrust, indicating a greater consensus around the responses for RW-VR.

\begin{table}
  \caption{Mean and standard deviations (SD) for trust scale for the
three prototype representations (max=7, min=1).}
  \label{tab:trust}
  \begin{tabular}{lcccc}
    \toprule
    &\textbf{RW-VR}&\textbf{CG-VR}&\textbf{RW-Video}&\textit{p-value}\\
    &(M / SD)&(M / SD)&(M / SD)\\
    \midrule
    \vspace{0.2cm}
    \textbf{Trust} & \textbf{4.98} / 0.75 & \textbf{4.64} / 1.25 & \textbf{4.37} / 1.29 & 0.55\\
    \textbf{Distrust} & \textbf{2.01} / 0.93 & \textbf{2.38} / 1.41 & \textbf{2.65} / 1.6 & 0.54\\
  \bottomrule
\end{tabular}
\end{table}

\subsubsection{Motivators of Trust}
\label{reasons_to_trust}
To better understand why participants generally trusted the vehicle (with low distrust across all three conditions), we coded the sections of the first interviews where we asked about reasons to trust. This allowed us to investigate the immediate responses given by the participants, considering which aspects could have positively influenced their trust. The codes and frequencies show that they primarily assessed trustworthiness based on the behaviour displayed by the vehicle itself, with similar frequencies verified among the three prototype representations. That is reasonable, given that the vehicle behaviour was identical across all representations. For example, participants reported that seeing the vehicle stopping for other pedestrians in the scenario \textit{`reinforced'} their trust (n=12, RW-VR=4, CG-VR=3, RW-Video=5). P21 mentioned that \textit{`after seeing [the vehicle] stop for someone, it wasn't too scary anymore'} (RW-VR). Others stated that seeing the vehicle \textit{`safely'} picking up another person in a previous scene (RW-VR=3) and then stopping in a safe distance from the participant (n=2, RW-VR=1, CG-VR=1) strengthened their sense of trust. The low speed of the vehicle was another reason that reinforced trust (n=6, RW-VR=3, CG-VR=1, RW-Video=2), among other contextual factors. For example, five participants stated that the light patterns communicated from the vehicle were the main driver for trusting it (RW-VR=1, CG-VR=3, RW-Video=1), while one participant (P3, RW-VR) stated that \textit{`the passengers in the car looked quite relaxed'}, which made her feel \textit{`more trusting'}. Finally, three participants related their trust to a general preexisting confidence in technology and autonomous systems (RW-VR=1, CG-VR=1, RW-Video=1).

\subsubsection{Comparisons Between Prototype Representations}
After experiencing all three prototype representations, roughly a quarter of the participants reported in the post-study interview that their subjective perception of trust towards the vehicle had not changed (n=11). These participants mentioned that the vehicle's driving behaviour in the presented scenarios was similar (e.g. in terms of speed, slowing down and communicating with pedestrians), so \textit{`it doesn't matter what kind of platform to use [for representation]'} (P32). However, more than a third of the participants stated that they trusted the AV more in the RW representations (n=15), with 11 of them explicitly reporting a higher perception of trust in RW-VR. Additionally, 4 participants stated having had less trust towards the AV in CG-VR, whereas 2 participants felt \textit{`more trustworthy'} (P11) or expressed to \textit{`feel more safe'} (P28) in CG-VR. Two participants stated that they experienced higher trust in both of the VR prototype representations, whereas one participant stated the opposite. The rest of the participants expressed difficulties to compare their perception of trust, feeling they have learned the interface after being exposed to the AV interactions multiple times (n=2). Others explicitly stated that their trust towards the situation changed, but that this did not influence their trust towards the car (n=2). 
\newline

To identify factors influencing participants' perception of trust, we included a question to that end in the post-study interview, aggregating responses into the high-level categories presented below: 

\begin{itemize}[label={}]

    \item (1) Spatial awareness: We found that participants' spatial awareness influenced their trust towards the AV to some degree (n=2). For example, P11 voiced the difficulty in assessing trust in RW-Video: \textit{`it didn't engage me enough to have any feelings about it, it was very distant’}. We further found that participants had better perception of space, distance and speed in the VR representations (n=6). Being immersed in VR, they \textit{`felt more’} (P27), had \textit{`a closer view of what could happen'} (P34) and noticed \textit{`more details’} (P31). Both distance between objects, and between participants and the vehicle, were better estimated in RW-VR and CG-VR as \textit{‘your whole body is within that environment, so you see the sizes of things,’} (P11) and \textit{‘everything was at a certain scale’} (P26). 
    These differences in spatial awareness prompted different emotional and behavioural reactions. For example, P27 became more concerned about the inattentive pedestrian: \textit{`When I watched the video, I thought that car can just gently bump into the pedestrian, it’s not a problem. When I saw it in VR, in 360-degree video, that would not have been a good idea'}. P26 grew more aware of the vehicle's trajectory: \textit{`I felt like where I was standing I was in its way and I wanted to step back out of [...] the path of the vehicle'}.
    
    \item (2) Realism of vehicle behaviour: Participants associated their higher level of trust towards the RW representations with a higher realism of the depicted driving behaviour (n=6). For example, P4 stated that the vehicle in CG-VR \textit{`felt more like fast and manic, and made [her] feel more manic too'}, which P27 related to the \textit{`sideways gliding'} of the vehicle. P10 referred to the \textit{`stability'} of the actual vehicle seen in RW-VR \textit{`that seemed a lot heavier [...] and stuck on the ground'}, and therefore \textit{`something [she] would jump on and trust'}. Other participants also mentioned that the stopping behaviour in CG-VR felt less realistic and thus less trustworthy (n=5). For example, P38 observed that \textit{`[the vehicle] stopped already at some distance [as] if it was by design to stop, not because it had seen the person there [through a sensor]'}. P22 mentioned in retrospect that for CG-VR she was not sure \textit{`if what was being depicted, was what the vehicle would be supposed to do'}, because \textit{`you can do all sort of things [in CG-VR]'}. Conversely, she also added that due to the lower realism she \textit{`allowed [the system] to be a bit wrong and still taking in what it said it would do'}. Participants reported that having seen in the RW representations that the vehicle can safely operate in the real world increased their level of trust (n=6). P1 explained, for example, that the authenticity of the RW representations \textit{`gives you a real behaviour of the code of the car'}, whereas P5 was impressed that \textit{`the vehicle is out there operating in the real world, [...] you know, it's not a science fiction movie - this is really happening'}. P23 added in this regard: \textit{`When it was just a simulation, it just feels like ``this is an idea but it’s not reality'', so I don’t think I would’ve had as much trust in it compared to the [RW-VR]'}.
    
    \item (3) Realism of people and the environment: We found that the different levels of realism in the depiction of people and the environment in the RW representations compared to CG-VR influenced the trust of participants towards the situation (n=7). Two female participants who experienced CG-VR as the initial representation stated in the first interview that they felt not very trustworthy towards the male characters in the scene. P14 mentioned that she was wondering \textit{`what [her] relationship to that [...] man was meant to be'} and that \textit{`it took a lot of [her] attention at the beginning'}. Similarly, P18 stated that \textit{`[she] was constantly looking at the guy next to [her]} and at some stage asked herself \textit{`what if I punch him?'}. She explained this reaction as a \textit{`fight or flight response'}, which was likely triggered by a slight uncanny valley affect and aggravated by the fact that the animated character did not respond in any way to the participant looking at the character. In the post-study interview, she referred back to this observation, commenting on a different emotional response when experiencing the RW-VR prototype representation: \textit{`I didn't feel like I wanted to hit the people because [in the RW-VR environment] they were understandable and they made sense'}. Similarly, P3 referred to the people in the RW-VR as (\textit{`look[ing] quite relaxed'}), which she linked to her perceived level of trust. The lower trust towards people in CG-VR was also influenced by the environment. For example, P14 mentioned that she instantly thought that \textit{`this is a place [she] would not stand as a woman to wait for a car'}, whereas in regards to the RW-VR prototype representation she mentioned that \textit{`this would be a situation I would be catching an Uber'}. The difference in the feelings reported towards people and the environment, however, was not matched by the perceptions of trust towards the vehicle. For example, P13 stated: \textit{`I trust more the situation [in RW-VR], but it doesn't affect my feeling to the car'}. Those participants also reported that trust towards the overall experience is more important for them. As explained by P18: \textit{`I trust the car, but I don't trust the other person [in the car]'}.
    
\end{itemize}

\subsection{User Experience (RQ3)}

\subsubsection{UEQ Questionnaire}
\begin{figure}
  \centering
  \includegraphics[width=0.95\linewidth]{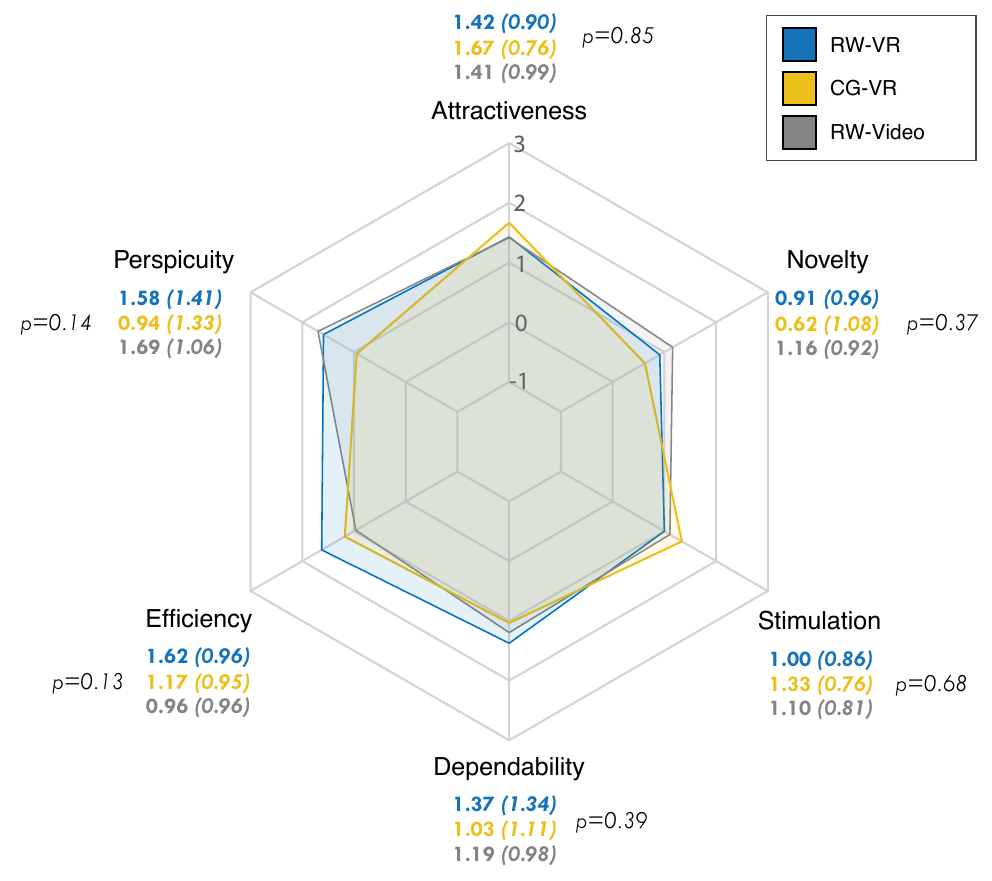}
  \caption{UX assessment of the eHMI across the three representations, based on the UEQ questionnaire
  \Description{Plotted graph of the UEQ questionnaire results, including results of the statistical analysis.}
  \cite{Laugwitz2008}.}
  \label{fig:ueq}
\end{figure}

Figure \ref{fig:ueq} shows the results of our descriptive data analysis of the UEQ scales across the three prototype representations. We were not able to find any significant differences between the different prototype representations. In other words, participants rated the UX of the eHMI similarly across CG-VR, RW-VR and RW-Video. 

\subsubsection{Differences in UX Feedback}
The analysis of the data from across both interviews revealed several aspects in relation to how participants assessed the UX of the eHMI.

\begin{itemize}[label={}]

\item (1) Comprehension: There were several lighting patterns introduced throughout the ride-sharing scenario. The analysis of the first interviews revealed that participants did not notice or pay attention to all of them. For example, the pattern when the vehicle was about to pull over was not often recalled upon when prompted in the post-study interview. We found the lack of attention towards the individual light patterns occurred more frequently in the VR representations (RW-VR=10, CG-VR=10) compared to the video representation (n=6). Similarly, the number of participants who correctly interpreted multiple light patterns was lower in CG-VR (n=6) and RW-VR (n=9) compared to RW-Video (n=12). Further, we only found explicit statements from participants in CG-VR that they were not able to understand the eHMI (n=3). Participants reasoned in this regard that in CG-VR they were distracted by the virtual depiction of people and environment (n=10). P10 stated that \textit{`everything looks kind of funny you tend to look around a lot’}. For P33, \textit{`the presence of that guy standing next to [her] waiting was really disturbing’}. Meanwhile, in RW-VR participants would get preoccupied by \textit{`other things'} (n=8). For example, P25 stated that \textit{`because it was too real, [she] focus[ed] on the surroundings, [...] and was looking at the sky as well for a moment’}. Another reason for distraction reported was the focus on finding the assigned car (n=3). As P19 expressed: \textit{`I was actually just trying to concentrate on which vehicle was mine. I didn’t think to look for any other additional information'}.

\item (2) Light colours: 
The participants were briefed before the experiment to wait for a car with the low-res light display showing a purple colour, while other cars had their light displays in blue. The analysis of the first interviews revealed that issues with distinguishing between the two colours was brought up by more participants in RW-VR (n=13) than in the other two representations (CG-VR=4, RW-Video=6). For example, they expressed confusion when a \textit{`seemingly purple’} car did not stop to pick them up. In this regard, they commented on the limitations of using colours for identifying a ride-sharing AV, which would be aggravated by scalability. For example, P19 stated: \textit{`If you’re talking about the use of colour as a vehicle distinguisher, it really depends on how concentrated the use of this type of vehicle is going to be. [...] imagine a crowd coming out of a sports stadium, there’s not going to be enough colours'}. Thus, participants experiencing the RW-VR prototype representation suggested the use of more unique combinations of colours, bespoke expressive light patterns, and high-resolution text displays. The reason why this was particularly critical in the RW-VR was that participants could not clearly distinguish between blue and purple from a distance. User feedback indicated a difference in display contrast between the two VR prototype representations. Half of the participants (n=21) mentioned that in CG-VR things looked clearer, \textit{`more sharp'} (P30), and \textit{`more vivid'} (P30), compared to the natural ambient light in the RW-VR (n=12). One participant pointed out that in regards to the CG-VR: \textit{`[...] in reality the lighting display will not be as noticeable and outstanding as in a [CG] simulation’} (P6).

\item (3) Contextual factors: A large number of participants also reflected in the first interviews on experiential aspects beyond the eHMI and the AV system, and thereby considered various situations they might face if they were to use the AV in real life. The majority of those statements were made by participants who experienced the VR representations (n=22, RW-VR=7, CG-VR=11, RW-Video=4). Nine participants expected extra cues, for example, via public displays in the environment or information displayed on their smartphones. P7 for example referred to displays at bus stops (\textit{`five minutes until your next bus’}) explaining that the information could enable her to \textit{`relax, sit down a bit, [...] have a drink, go to [have] a bathroom break’}. Participants also related to their personal habits and experiences (n=4). P15 thought that being able to identify the AV from afar based on the colour was really helpful, given that she might need time to \textit{`get off the phone and [...] grab [her] bags’}. P7 brought up a scenario in which she travelled with much luggage: \textit{`Is it actually going to start moving before I am able to get on safely? I’m halfway in and it [laughs] starts driving off?’}. P16 was anxious not knowing how long the AV would wait for her arrival as in reality she would \textit{`usually send the [driver] a message and ask them to wait’}.

\end{itemize}

\section{Discussion}
Our study results reveal a number of themes regarding the way participants responded to the three prototype representations and the feedback reported in the interviews. In this section, we discuss those themes and how they relate to our research questions regarding perceived sense of presence, trust and UX, followed by a series of design guidelines for context-based interface prototyping, and a reflection on study limitations.

\subsection{Effect of Prototype Representation on User Feedback}
\subsubsection{Sense of Presence} In terms of how the different prototype representations affected the sense of presence (RQ1), our results show, as expected, that the two VR representations (RW-VR and CG-VR) induced higher spatial presence and engagement compared to the video representation, echoing results from other VR simulator studies \cite{Dohyeon2020}. Interestingly, however, there was no significant difference in spatial presence between the two VR representations, despite various participants commenting on the unnatural impossibility of moving around in RW-VR. Regarding the naturalism of the scene (i.e. ecological validity), the quantitative and qualitative results both show that the RW-VR prototype representation more accurately depicted a real-world situation. Interviews confirmed that the lower perceived ecological validity of the scene in the CG-VR prototype representation was mainly induced by the diminished level of naturalism of the virtual characters and animated objects in the scene. Diminished immersion seemed also to have affected perceived ecological validity, which was ranked lower for RW-Video than for RW-VR, despite both displaying the same video material. The results from the ITC-SOPI questionnaire, emphasised through the semi-structured interviews, imply that RW-VR was best suited to induce a sense of sharing the same spatial context as the AVs in the scene. The high preference towards RW-VR (n=30) suggests that both immersion \textit{and} perceived naturalism are important design factors to consider when evaluating eHMI concepts with users.

\subsubsection{Trust}
In regards to how the prototype representations affected perceived user’s trust in the eHMI (RQ2), we can report that there were in fact three levels of trust simultaneously at play: (a) system trust, that is trust towards the vehicle and eHMI; 
(b) trust in the environment itself, including other people within it; and (c) trust in the real-world potential of the eHMI as a viable urban technology solution.

In terms of system trust, the quantitative results of our study indicated no significant difference in ratings between the three representations. Yet, qualitative feedback suggests that participants generally trusted the vehicle, and that they mainly derived their trust from the way the vehicle interacted with other pedestrians. 
Given that this behaviour (e.g. giving way to a pedestrian) was identical in all three prototype representations, this explains why there were no significant differences in the perception of trust towards the AV. 
Interestingly, in the post-study interviews, more than a third of the participants reported that they would trust the vehicle more in the RW representations. However, the detailed analysis revealed that participants' assessment of trust is based not just on the AV itself but also determined by trust towards the overall experience.

The increased sense of presence provided by the VR representations led participants to express feedback directed at particular elements of the environment (e.g. texture of trees) and, crucially, at other human beings in the scene, who they more tangibly \textit{felt} to be sharing the experience with. That is relevant, as the VR representations seemed to elicit a feedback loop between participants and the environment, prompting varying feelings of relatedness to strangers around them, and causing them to consider different behaviour in response to the different levels of realism provided by the prototype representation. Female participants feeling unsafe in the presence of a seemingly unresponsive (CG simulated) male character near to them, in a situation they felt having little control over, are an insightful example of how lower realism can affect the trust in the environment negatively. This is an important observation as it highlights that when assessing trust towards an AV prototype in a simulated environment, other factors might be at play that influence participants' responses. 

The realism of the experience offered by the RW representations, particularly the RW-VR, also seemed to boost participants' trust in the real-world viability of the eHMI in urban spaces. The direct experience of the technology contextualised to a real street and surrounded by real people conveyed to participants a sense of \textit{'this already [being] reality, not fiction'}, prompting them to reflect upon practical aspects such as safety (their own and others'), ambience (emotional cues given by other people in the scene) and the autonomous nature of the technology (more clearly decoupled from the surrounding environment, in comparison to the CG-VR representation).

\subsubsection{User Experience}
Regarding RQ3, the quantitative results of our study show that there is no significant difference in UEQ ratings between the three prototype representations. However, there are some tendencies in the ratings that are supported by the qualitative data. For example, attractiveness and stimulation is rated slightly higher in CG-VR, which also relates to the increased colour contrast participants reported on, and therefore offering a `cleaner' depiction of the low-res lighting interface. Further, higher ratings for perspicuity were matched by participant comments revealing that they found it easier to comprehend the lighting patterns in the RW representations. This can be linked to the fact that participants reported being distracted by various other aspects in CG-VR (such as the texture of trees). Interestingly, in RW-Video, participants noticed and were able to reflect on more of the eHMI light patterns in the subsequent interviews. This contradicts previous research which reported better memory assessment in immersive experiences \cite{Ventura2019}, and indicates that the VR experience itself - although being more vividly - can distract from the assessment of singular user interface elements. Participants confirmed this observation in the post-study interviews, for example, stating that they were more concerned about \textit{`finding their car'}. On the other hand, the VR prototype representations allowed them to \textit{`understand the user experience more holistically'} (P14), which also led to more detailed feedback on aspects beyond the eHMI.

\subsection{Guidelines for Prototyping and Evaluating Context-Based Interfaces}
Based on our comparative study involving a shared AV scenario and the evaluation of trust and UX towards a custom-designed eHMI, we propose a series of preliminary guidelines for prototyping and evaluating context-based interfaces for autonomous systems through simulations and videos.

\subsubsection{Choosing a Simulation Platform and Representation.}
The choice of simulation platform (e.g. VR or video) and prototype representation (e.g. CG or RW) depends on the specific questions that the evaluation seeks to address. 

\begin{itemize}[label={}]

\item \textbf{GL1 - Use non-immersive prototypes for focused interface evaluations:} We found the assessment of trust towards AVs in a simulated scenario to be heavily based on how the vehicle interacted with other pedestrians. These interactions between autonomous systems and other people sharing the same urban environment can easily be captured in video prototype representations, eliminating the need for a costly VR setup and supporting online evaluation studies \cite{Dey2020}. 
Indeed, we found that participants were able to remember and comment on the light patterns used in our eHMI better in RW-Video than the VR representations.

\item \textbf{GL2 - Use immersive prototypes for holistic assessment and evaluation of contextual aspects:} We learnt that the VR representations allowed for a more holistic assessment of the user's relationship with the eHMI in the simulated urban environment, due to increased spatial awareness and stronger sense of being actively present in the scene. We therefore propose that VR representations (RW and CG) are better suited when seeking user feedback on not only the interface but how the interface influences the user's experiential and perceptual aspects within a particular context.

\item \textbf{GL3 - Use real-world representations to increase familiarity and assess overall trust:} Previous work on driving simulators stressed that familiarity with the environment in real-world videos increases feeling of safety and leads to richer feedback \cite{Gerber2019}. High realism and influence of environmental factors as well as social interactions between multiple people sharing an urban space with an eHMI were also deemed important by our participants. RW-VR, thus, is especially well-suited for capturing the more nuanced aspects of trust beyond the system itself (linked to the complex and dynamic context within which the system operates).

\item \textbf{GL4 - Use real-world representations to uncover interface anomalies under more natural conditions:} We received more responses on potential interface anomalies (i.e. use of colour to encode information) and potential alternatives (i.e. text displays) in the real-world representations due to the more natural ambient lighting and lower contrast compared to the CG-VR. Therefore, we conclude that real-world representations might be better suited to evaluate the viability of visual-based interface design proposals.

\end{itemize}

\subsubsection{Composition of Scenes.} Prototyping and evaluating context-based interfaces within a simulated or captured real-world urban environment comes with a range of challenges and confounding factors compared to decontextualised evaluation setups (as used e.g. in \cite{Dey2020}). Scenes should be carefully composed and their composition has implications on perceived trust as well as keeping participant engagement high.

\begin{itemize}[label={}]

\item \textbf{GL5 - Stage interactions with context-based interfaces:} Our findings show that interactions between other pedestrians and the eHMI mainly contributed to the assessment of trust towards the AV. The high number of responses on those interactions further indicates that staged interactions increase memory assessment which can in turn lead to richer feedback in post-experience interviews. Additionally, staged interactions might prevent survey fatigue which is in particular important in RW-VR with the participant's own interaction radius being limited.

\item \textbf{GL6 - Consider effect of environment and people:} Our qualitative data showed that aspects beyond the system influenced participants' perception of trust and user experience. Indeed some participants deemed those aspects, such as with whom they would be sharing an AV and waiting for the vehicle in a dark, empty location, as more critical than the vehicle itself. We therefore conclude that it's crucial to consider the effect of surrounding entities, and in turn stress for the importance of contextualised simulation setups for a more holistic evaluation of interactions with autonomous systems.

\item \textbf{GL7 - Carefully consider camera position and constrain movement in RW-VR:} Due to the lack of freedom to move in RW-VR, the camera position for recording has to be carefully chosen. In our specific context, it was important that the camera was not positioned in the vehicle's potential trajectory, while still warranting a good viewing angle to observe the interactions. Positioning people next to the camera or recording in a physically constraint environment can further help to create a visual bounding box to deter participant's urge to move within the RW-VR environment.

\end{itemize}

\balance



\subsubsection{Designing CG-VR Prototypes.} Finally, our findings suggests considerations to be made for the specific case of prototyping context-based interfaces through CG-VR representations.

\begin{itemize}[label={}]

\item \textbf{GL8 - Avoid virtual avatars in intimate or personal proxemic zones of participants:} Our study results indicate that computer-generated context-based interface evaluations, which require simulated avatars to interact with autonomous systems, can be affected by the uncanny valley phenomenon \cite{Latoschik2017}. This not only leads to decreased perception of realism, but also in decreased trust towards the overall experience and feelings of distraction compromising the assessment of the actual prototype. Based on that observation, we recommend that virtual avatars, if possible, should not be placed in the intimate or personal proxemic zones of participants.

\item \textbf{GL9 - Avoid unnecessary details to prevent distraction by imperfections in CG representations:} Aiming for an accurate copy of the RW-VR source in CG-VR, our 3D designer carefully crafted fine details, such as tree textures and animation of leafs to simulate wind. However, our participants reported those details as having been distracting and emphasising imperfections in CG-VR. Due to the still apparent lack of realism in CG representations and given the often limited budget for research prototypes, we therefore recommend to limit unnecessary details (and in particular animations) concerning the surrounding environment.

\end{itemize}

\subsection{Limitations and Future Work}
The presented study has some limitations that we would like to acknowledge. 
To minimise the learning effect and transfer across conditions, we opted for a between-subject design, which, however, comes with the limitation that less data points per participant are taken. Although the number of participants is in the range of other similar studies (cf. \cite{Pettersson2019}), we acknowledge that the sample size for the quantitative data analysis is rather small. Yet, we would also argue that the quantitative data is only part of the broader scope of participant data collected, and feeds into the additional analysis of qualitative data from 11.5 hours of interviews. 

The novelty effect inherent to emerging technologies, such as VR, and participants' previous experience with VR, might also have had an impact on the study results. We tried to address this as much as possible by counterbalancing previous experiences in VR in our study design. We further investigated the collected data for differences between participants linked to their previous experience. We found that participants with no previous VR experience were impressed by the high realism and immersion of the CG-VR when experiencing this representation as the first condition. After subsequently having experienced the RW-VR, they stated that they would have assessed their sense of presence in CG-VR differently. The novelty effect in our study may be further affected by the fact that none of the participants (including those with previous VR experience) had experienced 360-degree VR before.

Previous experimental research studies with autonomous driving simulators have acknowledged the limitations of measuring trust based on post-experiment questionnaires and interviews \cite{Frison2019, Hollaender2019}. They also refer to previous work in human-robot interaction, which highlights that a widely accepted definition of trust is missing \cite{Lewis2018}.
While acknowledging this as a limitation, we also want to emphasise the exploratory findings we gained through interviews, for example, showing that participants assess trust towards various entities in a VR simulation.
Furthermore, when using RW representations, participants seem to factor in potential real-life consequences in their perception of trust, resulting in increased feelings of alertness and awareness of the environment. We therefore posit that these findings offer new insights on the multifaceted and complex aspects of measuring trust towards autonomous systems in VR.

Further, given that we only conducted a single user study, we see our findings and synthesised guidelines as preliminary and not set in stone, indicating areas of future work and requiring more focused investigations: for example, in regards to the use of avatars in CG-VR, future research should investigate the effects of low-realism or abstract representations of avatars on user feedback during context-based interface evaluations. Plus, as suggested by a participant, it might be helpful to allow users to visualise their own body parts in the same visual style as the avatars in the scene \cite{Rebelo2012, Lawson2016}, so that they can better relate their own virtual self to the simulated characters. Another open challenge for evaluating context-based interfaces is to find a sweet spot between preventing survey fatigue \emph{and} offering sufficient time to experience the prototype. A potential solution for keeping participants engaged also during longer scenarios in VR could be to enable them to interact with a smartphone in meaningful ways that support the scenario. 

\section{Conclusion}
To sum up, the advent of AVs brings new challenges into the domain of interaction design, such as prototyping and evaluating context-based interfaces (e.g. eHMIs). At the same time, technological advances in immersive video capturing and VR hardware offers designers and researchers a wider range of possible prototyping representations and platforms to choose from. 
By systematically studying the effect of prototyping representations on study results, our paper adds to previous work on virtual field studies \cite{Maekela2020} and context-based interface prototyping \cite{Flohr2020}.

\begin{acks}
This research was supported partially by the Sydney Institute for Robotics and Intelligent Systems (SIRIS) and ARC Discovery Project DP200102604 Trust and Safety in Autonomous Mobility Systems: A Human-centred Approach. The authors acknowledge the statistical assistance of Kathrin Schemann of the Sydney Informatics Hub, a Core Research Facility of the University of Sydney. We thank all the participants for taking part in this research. We also thank the anonymous CHI’21 reviewers and ACs for their constructive feedback and suggestions how to make this contribution stronger.
\end{acks}


\balance{}
\balance{}
\bibliographystyle{ACM-Reference-Format}
\bibliography{sample-base}

\end{document}